\documentclass[preprintnumbers,superscriptaddress,showpacs,pra]{revtex4}
\usepackage{amssymb,amsmath}
\usepackage{graphicx}
\usepackage{amsmath}
\usepackage{amsfonts}

\input{tcilatex}
\begin{document}

\preprint{}
\title[]{Quantum motion on a torus as a submanifold problem in a generalized
Dirac's theory of second-class constraints}
\author{D. M. Xun}
\affiliation{School for Theoretical Physics, and Department of Applied Physics, Hunan
University, Changsha, 410082, China}
\author{Q. H. Liu}
\email{quanhuiliu@gmail.com}
\affiliation{School for Theoretical Physics, and Department of Applied Physics, Hunan
University, Changsha, 410082, China}
\author{X. M. Zhu}
\affiliation{School of Science, Xidian University, Xi'an 710071, China}
\date{\today }

\begin{abstract}
A generalization of the Dirac's canonical quantization theory for a system
with second-class constraints is proposed as the fundamental commutation
relations that are constituted by all commutators between positions, momenta
and Hamiltonian so they are simultaneously quantized in a self-consistent
manner, rather than by those between merely positions and momenta so the
theory either contains redundant freedoms or conflicts with experiments. The
application of the generalized theory to quantum motion on a torus leads to
two remarkable results: i) The theory formulated purely on the torus, i.e.,
based on the so-called the purely intrinsic geometry, conflicts with itself.
So it provides an explanation why an intrinsic examination of quantum motion
on torus within the Schr\"{o}dinger formalism is improper. ii) An extrinsic
examination of the torus as a submanifold in three dimensional flat space
turns out to be self-consistent and the resultant momenta and Hamiltonian
are satisfactory all around.
\end{abstract}

\maketitle

\section{Introduction}

The embedding problem of quantum motion of a particle on a two-dimensional
curved surface $\Sigma ^{2}$ in the flat space $R^{3}$ has attracted much
attention, including theoretical explorations \cite%
{jk,dacosta,CB,FC,liu07,liu11,japan1990,japan1992,japan1993}\ and
experimental investigations \cite{Szameit,onoe}. Fundamentally, there are
two formalisms to investigate the quantum motion on $\Sigma ^{2}$. One is
within the Schr\"{o}dinger formalism that needs a wave function and another
is within the Dirac one that purely deals with operators, but they usually
give different predictions. In this section, we will mainly review these two
formalisms, and present a generalization of the Dirac's canonical
quantization theory for a system of the second-class constraints.

\subsection{Schr\"{o}dinger and Dirac formalism:\ discrepancies in curvature
dependent quantum potentials}

By the \textit{Schr\"{o}dinger formalism} we mean that the Schr\"{o}dinger
equation is first formulated in $R^{3}$, actually in a curved shell of an
equal and finite thickness $\delta $ whose intermediate surface coincides
with the prescribed one $\Sigma ^{2}$ (or equivalently, the particle moves
within the range of the same width $\delta $ due to a confining potential
around the surface), and an effective Schr\"{o}dinger equation on the curved
surface $\Sigma ^{2}$ is then derived by taking the squeezing limit $\delta
\rightarrow 0$ to confine the particle to the $\Sigma ^{2}$ \cite%
{jk,dacosta,CB,liu11}. It leads to a unique form of the so-called geometric
potential \cite{Szameit,liu11} 
\begin{equation}
V_{g}=-\frac{\hbar ^{2}}{2m}\left( M^{2}-K\right)  \label{gp}
\end{equation}%
that depends on both the mean and\ the gaussian curvature $M$ and $K$ which
are, respectively, the extrinsic and the intrinsic curvature. This amounts
to an extrinsic examination of the quantum motion on $\Sigma ^{2}$ within
the Schr\"{o}dinger formalism. The potential (\ref{gp}) has been
experimentally confirmed \cite{Szameit,onoe}. To note that the extrinsic
curvature $M$ is a geometric consequence of embedding the system on $\Sigma
^{2}$ in $R^{3}$ and is inaccessible with purely intrinsic description.
However, for this formalism, we do not know why the Schr\"{o}dinger equation
can not be entirely formulated on $\Sigma ^{2}$ without considering any
embedding. We are familiar with a fact an intrinsic examination of the
quantum motion on $\Sigma ^{2}$ within the Schr\"{o}dinger formalism that
predicts no curvature dependent quantum potential, which is contrary to the
experiments \cite{Szameit,onoe}.

By the \textit{Dirac formalism} we mean to use the Dirac's canonical
quantization theory on systems with the second-class constraints \cite%
{dirac1,dirac2}, with an understanding that Dirac formalism can also be
applied to the system\textit{\ }that is considered either within purely
intrinsic geometry on $\Sigma ^{2}$ or as a submanifold in $R^{3}$,
predicting a curvature dependent potential $V_{D}$ with two real parameters $%
\alpha $ and $\beta $ \cite{japan1990,japan1992}, 
\begin{equation}
V_{D}=-\frac{\hbar ^{2}}{2m}\left( \alpha M^{2}-\beta K\right) .  \label{vd}
\end{equation}%
This form of the potential (\ref{vd}) can also be easily constructed by
dimensional analysis for two geometric invariants $M$ and $K$ have dimension
of \textit{length}$^{-1}$ and \textit{length}$^{-2}$, respectively. In
comparison with the Schr\"{o}dinger formalism, we have one more unknown
associated with the Dirac one, that is, once taking the $\Sigma ^{2}$ as a
submanifold in $R^{3}$ we do not know what form of the potential can be
singled out among a family of it (\ref{vd}). However, Schr\"{o}dinger's
theory gives an unambiguous choice with $\alpha =\beta =1$ \cite%
{jk,dacosta,FC,liu11}.

So far, we find that both formalisms suffer from shortcomings. Since the
extrinsic examination of the torus within the Schr\"{o}dinger formalism has
experimental supports, an immediate question is whether there is a possible
theoretical framework from which we can fix the parameters $\alpha $ and $%
\beta $ within a possibly generalized Dirac's theory, so rendering it
compatible with Schr\"{o}dinger's and also the experimental results. This
question will be partially answered in this paper.

\subsection{Schr\"{o}dinger and Dirac formalism:\ discrepancies in momentum
operators}

In addition to the unique form of the geometric potential $V_{g}=-\hbar
^{2}\left( M^{2}-K\right) /2m$, Schr\"{o}dinger's theory also leads to a
unique definition of the geometric momentum $\mathbf{p}$ \cite{liu07,liu11}, 
\begin{equation}
\mathbf{p}=-i\hbar (\mathbf{r}^{\mu }\partial _{\mu }+M\mathbf{n}),
\label{gm}
\end{equation}%
where $\mathbf{r=(}x(x^{1},x^{2}),y(x^{1},x^{2}),z(x^{1},x^{2})\mathbf{)}$
is the position vector in $R^{3}$ on the surface $\Sigma ^{2}$ whose local
coordinates are $x^{\mu }\equiv (x^{1},x^{2})$ and $\mathbf{r}^{\mu }=g^{\mu
\nu }\mathbf{r}_{\nu }=g^{\mu \nu }\partial \mathbf{r/}x^{\nu }$, and at
this point $\mathbf{r}$, $\mathbf{n=(}n_{x},n_{y},n_{z}\mathbf{)}$ denotes
the normal and $M\mathbf{n}$ symbolizes the mean curvature vector field,
another geometric invariant. Throughout the paper, the Einstein summation
convention over repeated indices is used.\ 

However, the present formulation of Dirac's theory opens a wide door to
permit various definitions of the\ generalized momenta, including i) the
well-known generalized ones $p_{\mu }=-i\hbar (\partial _{\mu }+\Gamma _{\mu
}/2)$ which satisfy quantum commutator $[x^{\nu },p_{\mu }]=i\hbar \delta
_{\mu }^{\nu }$, where $\Gamma _{\mu }$ is the once-contracted Christoffel
symbol $\Gamma _{\mu \nu }^{\sigma }$ constructed with Riemannian metric $%
g^{\mu \nu }$ \cite{japan1990}, where greek letters $\mu $, $\nu $, $\sigma $%
, etc. run between $1$ to $2$, and ii) geometric momentum (\ref{gm}), and
etc. \cite{liu11,japan1992}. \textit{It is very important to note that in
the extrinsic examination of quantum motion on }$\Sigma ^{2}$\textit{\ in }$%
R^{3}$\textit{, the local coordinates }$x^{\mu }\equiv (x^{1},x^{2})$\textit{%
\ are no longer position operators but parameters, and the position
operators are }$\mathbf{r=(}x(x^{1},x^{2}),y(x^{1},x^{2}),z(x^{1},x^{2})%
\mathbf{)}$\textit{.}

A framework based on the purely intrinsic geometry implies that every
quantity solely relies on the Riemannian metric $g^{\mu \nu }$ and its
various constructions such as Christoffel symbol $\Gamma _{\mu \nu }^{\sigma
}$ and the gaussian curvature $K$. Consequently, neither momentum nor
Hamiltonian in quantum mechanics depends on the extrinsic curvature. When
the curvature dependent potential with$\ $(\ref{vd}) $\alpha \neq 0$ and
geometric momentum (\ref{gm}) appear in a formulation of quantum mechanics
for a system on $\Sigma ^{2}$, we in fact take the system under study to be
embedded in $R^{3}$, which is beyond the purely intrinsic geometry.

\subsection{A generalization Dirac's theory for a system of the second-class
constraints}

We are deeply impressed by the very success of Schr\"{o}dinger's theory that
produces unique result of the geometric potential (\ref{vd})\ and momentum (%
\ref{gm}), and also by the disturbing arbitrariness associated with Dirac's
theory of the second-class constraints. As we know, Dirac's theory
postulates that a quantum commutator $[A,B]$ of two variables $A$ and $B$ in
quantum mechanics is achieved by direct correspondence of the Dirac's
brackets $\{A{,B\}}_{D}$ as $\{A{,B\}}_{D}\rightarrow \lbrack A,B]\ $which
is defined by $[A,B]=i\hbar O(\{A{,B\}}_{D})$ where $O(F)$ is used to
emphasize the operator form of the classical quantity $F$ in order to avoid
possible confusion. When all constraints are removed, the Dirac bracket $\{A{%
,B\}}_{D}$ assumes its usual\textit{\ }form, the Poisson\textit{\ }bracket $%
\{A{,B\}}$. However, Dirac himself states that \textit{fundamental
commutation relations involve only those between canonical positions }$x_{{i}%
}$\textit{\ and canonical momenta} ${p}_{{i}}$\textit{\ }\cite{dirac1,dirac2}%
.

One can ask a curious question: when there is no constraint, why there is no
such a \textit{fundamental} canonical quantization rule between $f$ $(=x_{{i}%
}$, $p_{{i}})$ and the Hamiltonian $H$ as $[f,H]=i\hbar O(\{f,H\})$? This is
because the direct quantization $[f,H]=i\hbar O(\{f,H\})$ might be
redundant, or meaningless, or practically useless, etc. For instance, when
the system has a classical analogue, the Hamiltonian is the same function of
the positions and momenta in the quantum theory as in the classical theory,
provided that the Cartesian system of axes is used \cite%
{dirac2,dirac3,Greiner}. In this case the rule $[f,H]=i\hbar O(\{f,H\})$
turns out to be redundant. When a quantum Hamiltonian has no classical
analogy, the canonical quantization rule $[f,H]=i\hbar O(\{f,H\})$ is
meaningless. In many other cases, e.g., to quantize a classical Hamiltonian $%
H=\gamma x^{3}p^{3}$ with $\gamma $ being a real parameter, the rule should
be imposed but is practically useless. Thus, it appears unacceptable to
include the\ canonical quantization rule $[f,H]=i\hbar O(\{f,H\})$ as a
fundamental element of a theory.

For systems with the second-class constraints, the situation is totally
different. Discrepancies between either curvature dependent quantum
potentials or momentum operators present when different formalisms, or
different geometric points, are utilized. It strongly implies that, while
the quantization of the system is performed, the proper operator form of
positions, momenta and Hamiltonian are simultaneously determined in a
self-consistent way. Therefore we have attempted to generalize the Dirac's
theory so as to add $[f,H]$ into the category of the \textit{fundamental
commutation relations} which should also be directly achieved via following
quantization rule \cite{liu11}, 
\begin{equation}
\lbrack f,H]=i\hbar O(\{f,H\}_{D}),\text{ }f=x_{i}\text{ and }p_{j}.
\label{generalized}
\end{equation}%
In rest part of the paper, the convention $O(F)=F$ in quantum mechanics
assumes without no longer emphasizing it an operator with the symbol $O$.
These commutation relations (\ref{generalized}) may not be applicable when
the system has no constraint. So we would like to call them the second
category of fundamental\textit{\ }ones \cite{liu11}, whereas the existing
ones between positions and momenta, the first.

This generalized Dirac's theory reproduces the usual form for the system
that has a classical analogue but has not a constraint, together with the
necessary utilization of the Cartesian system of axes, therefore enriches
the Dirac formalism of quantum mechanics. We will call it the \textit{%
general theory of the canonical quantization} (GTCQ).

\subsection{Purpose and organization of the paper}

As\ an application of the GTCQ to quantum motion on a sphere \cite{liu11},
we find that, on one hand, an attempt of trying a proper description within
the purely intrinsic geometry proves problematic, and one the other hand, an
account of embedding the sphere in three-dimensional space is very coherent.
Notice that the classification theorem for compact surfaces states that \cite%
{class}, every compact orientable surface is homeomorphic either to a sphere
or to a connected sum of tori, implying that if there is any difficulty
associated with quantum mechanics for a particle constrained on a sphere or
a torus, enormous theoretical problems would arise from dealing with an
arbitrary two-dimensional curved surface in quantum mechanics. It forms one
of the reasons that the sphere \cite{liu11}\ and the torus \cite%
{torus1,torus2,torus3,torus4} are used to test various theories. The main
purpose of the present study is to\ take the torus to show that Dirac
formalism is complementary to the Schr\"{o}dinger one. The former eliminates
the purely intrinsic description, and the latter gives the unique form of
the geometric potential, while both define the identical form of the
geometric momentum.

This paper is organized as follows. In following section II, we present the
GTCQ for quantum motion on the torus within purely intrinsic geometry.
Results show that the theory can never be consistently set up. In section
III, we revisit the same problem as a submanifold in flat space $R^{3}$ with
the GTCQ. Results show that the theory turns out to be self-consistent all
around, and the obtained geometric momentum (\ref{gm}) and potential (\ref%
{gp}) are also satisfactory. Section IV briefly remarks and concludes this
study.

\section{GTCQ for a torus within intrinsic geometry}

The toroidal surface is with two local coordinates $\theta \in \lbrack
0,2\pi ),\varphi \in \lbrack 0,2\pi )$%
\begin{equation}
\mathbf{r}=((a+r\sin \theta )\cos \varphi ,(a+r\sin \theta )\sin \varphi
,r\cos \theta ),\text{ }a>r\neq 0,  \label{rr}
\end{equation}%
where $\varphi $ is the azimuthal angle and $\theta $ the polar angle, and $%
a $ and $r$ are the outer and inner radii of the torus, respectively. The
constraint is $r=b\neq 0$. In this section, we will first give the classical
mechanics for motion on the torus, and then turn into the Dirac formalism of
quantum mechanics. In classical mechanics, the theory appears nothing
surprising, but after transition to quantum mechanics, it becomes
contradictory to itself.

\subsection{Classical mechanical treatment}

The Lagrangian $L$ in the toric coordinate system is, 
\begin{equation}
L=\frac{m}{2}(\dot{r}^{2}+r^{2}\dot{\theta}^{2}+(a+r\sin \theta )^{2}\dot{%
\varphi}^{2})-\lambda (r-b),  \label{lag}
\end{equation}%
where $\lambda $ is the Lagrangian multiplier enforcing the constrained of
motion on the surface. The Lagrangian is singular because it does not
contain the "velocity" $\dot{\lambda}$. Hence we need the Dirac formalism of
the classical mechanics for a system with the second-class constraints,
which gives the canonical momenta conjugate to $r,\theta ,\varphi $ and $%
\lambda $ in the following,%
\begin{eqnarray}
p_{r} &=&\frac{\partial L}{\partial \dot{r}}=m\dot{r}, \\
p_{\theta } &=&\frac{\partial L}{\partial \dot{\theta}}=mr^{2}\dot{\theta},
\\
p_{\varphi } &=&\frac{\partial L}{\partial \dot{\varphi}}=m(a+r\sin \theta
)^{2}\dot{\varphi}, \\
p_{\lambda } &=&\frac{\partial L}{\partial \dot{\lambda}}=0.  \label{plamb}
\end{eqnarray}%
Eq. (\ref{plamb}) represents the primary constraint:%
\begin{equation}
\varphi _{1}\equiv p_{\lambda }\approx 0,  \label{prim}
\end{equation}%
hereafter symbol "$\approx $" implies a weak equality \cite{dirac2}. After
all calculations are finished, the weak equality takes back the strong one.
By the Legendre transformation, the primary Hamiltonian $H_{p}$ is \cite%
{dirac2},%
\begin{equation}
H_{p}=\frac{1}{2m}(p_{r}^{2}+\frac{p_{\theta }^{2}}{r^{2}}+\frac{p_{\varphi
}^{2}}{(a+r\sin \theta )^{2}})+\lambda \left( r-b\right) +\dot{\lambda}%
p_{\lambda },  \label{hami}
\end{equation}%
where $\dot{\lambda}$ is also a Lagrangian multiplier guaranteeing that this
Hamiltonian is defined on the symplectic manifold. The secondary constraints
(not confusing with second-class constraints) are generated successively,
then determined by\ the conservation condition \cite{dirac2},%
\begin{equation}
\varphi _{i+1}\equiv \left\{ \varphi _{i},H_{p}\right\} \approx 0,\text{\ }%
(i=1,2,....),
\end{equation}%
where $\left\{ f,g\right\} $ is the Poisson bracket with $%
q_{1}=r,q_{2}=\theta ,q_{3}=\varphi $, and $p_{1}=p_{r},p_{2}=p_{\theta
},p_{3}=p_{\varphi }$, 
\begin{equation}
\left\{ f,g\right\} \equiv \frac{\partial f}{\partial q_{k}}\frac{\partial g%
}{\partial p_{k}}+\frac{\partial f}{\partial \lambda }\frac{\partial g}{%
\partial p_{\lambda }}-(\frac{\partial f}{\partial p_{k}}\frac{\partial g}{%
\partial q_{k}}+\frac{\partial f}{\partial p_{\lambda }}\frac{\partial g}{%
\partial \lambda }).  \label{possi}
\end{equation}%
The complete set of the secondary constraints is, 
\begin{eqnarray}
\varphi _{2} &\equiv &\left\{ \varphi _{1},H_{p}\right\} =-(r-b)\approx 0,
\label{db1} \\
\varphi _{3} &\equiv &\left\{ \varphi _{2},H_{p}\right\} =-\frac{p_{r}}{m}%
\approx 0,  \label{db2} \\
\varphi _{4} &\equiv &\left\{ \varphi _{3},H_{p}\right\} =\frac{\lambda }{m}-%
\frac{1}{m^{2}}(\frac{p_{\theta }^{2}}{r^{3}}+\frac{p_{\varphi }^{2}\sin
\theta }{(a+r\sin \theta )^{3}})\approx 0,  \label{thi} \\
\varphi _{5} &\equiv &\left\{ \varphi _{4},H_{p}\right\} =\frac{\dot{\lambda}%
}{m}-\frac{3ap_{\theta }p_{\varphi }^{2}\cos \theta }{m^{3}r^{2}(a+r\sin
\theta )^{4}}\approx 0.  \label{for}
\end{eqnarray}%
Eqs. (\ref{db1}) and (\ref{db2}) show, respectively, that on the surface of
torus $r=b$, no motion along the normal direction is possible $p_{r}=0$,
while Eqs. (\ref{thi}) and (\ref{for}) determine, respectively, the
Lagrangian multipliers $\lambda $ and $\dot{\lambda}$.

The Dirac bracket instead of the Poisson one for two variables $A$ and $B$
is defined by,%
\begin{equation}
\left\{ A,B\right\} _{D}\equiv \left\{ A,B\right\} -\left\{ A,\varphi
_{u}\right\} C_{uv}^{-1}\left\{ \varphi _{v},B\right\} ,
\end{equation}%
where the $4\times 4$ matrix $C\equiv \left\{ C_{uv}\right\} $ whose
elements are defined by $C_{uv}\equiv \left\{ \varphi _{u},\varphi
_{v}\right\} $ with $u,v=1,2,3,4$ from Eqs. (\ref{prim}) and (\ref{db1})-(%
\ref{thi}). The inverse matrix $C^{-1}$ is,%
\begin{equation}
C^{-1}=\left\{ 
\begin{array}{cccc}
0 & C_{12}^{-1} & 0 & m \\ 
-C_{12}^{-1} & 0 & -m & 0 \\ 
0 & m & 0 & 0 \\ 
-m & 0 & 0 & 0%
\end{array}%
\right\} ,
\end{equation}%
where%
\begin{equation}
C_{12}^{-1}=\frac{3}{m}\left( \frac{p_{\theta }^{2}}{b^{4}}+\frac{p_{\varphi
}^{2}\sin ^{2}\theta }{\left( a+b\sin \theta \right) ^{4}}\right) .
\end{equation}%
Thus, the generalized positions $q^{\mu }$ $(=\theta ,\varphi )$ and momenta 
$p_{\mu }$ satisfy the following Dirac brackets,%
\begin{equation}
\{q^{\mu },q^{\nu }\}_{D}=0,\text{ }\{p_{\mu },p_{\nu }\}_{D}=0,\text{ }%
\{q^{\mu },p_{\nu }\}_{D}=\delta _{\nu }^{\mu }.  \label{xp1}
\end{equation}%
By use of the equation of motion,%
\begin{equation}
\dot{f}=\left\{ f,H_{c}\right\} _{D},
\end{equation}%
we obtain those for the positions $\theta $, $\varphi $ and the momenta $%
p_{\theta }$, $p_{\varphi }$, respectively,%
\begin{eqnarray}
\dot{\theta} &\equiv &\left\{ \theta ,H_{c}\right\} _{D}=\frac{p_{\theta }}{%
mb^{2}},\text{\ \ }\dot{\varphi}\equiv \left\{ \varphi ,H_{c}\right\} _{D}=%
\frac{p_{\varphi }}{m(a+b\sin \theta )^{2}},  \label{xh} \\
\dot{p}_{\theta } &\equiv &\left\{ p_{\theta },H_{c}\right\} _{D}=\frac{%
b\cos \theta p_{\varphi }^{2}}{m(a+b\sin \theta )^{3}},\text{ \ }\dot{p}%
_{\varphi }\equiv \left\{ p_{\varphi },H_{c}\right\} _{D}=0.  \label{ph}
\end{eqnarray}%
In these calculations (\ref{xh}) and (\ref{ph}), we in fact need only the
usual form of Hamiltonian, $H_{p}\rightarrow H_{c}$,%
\begin{equation}
H_{c}=\frac{1}{2m}\left( \frac{p_{\theta }^{2}}{b^{2}}+\frac{p_{\varphi }^{2}%
}{\left( a+b\sin \theta \right) ^{2}}\right) .
\end{equation}

So far, the classical mechanics for the motion on the torus is complete and
coherent in itself.\ 

\subsection{Quantum mechanical treatment}

In quantum mechanics, we assume that the Hamiltonian takes the following
general form,%
\begin{eqnarray}
H &=&-\frac{\hbar ^{2}}{2m}\left[ \nabla ^{2}+\left( \alpha M^{2}-\beta
K\right) \right]  \notag \\
&=&-\frac{\hbar ^{2}}{2m}\left[ \frac{1}{b^{2}}\frac{\partial ^{2}}{\partial
\theta ^{2}}+\frac{\cos \theta }{b\left( a+b\sin \theta \right) }\frac{%
\partial }{\partial \theta }+\frac{1}{\left( a+b\sin \theta \right) ^{2}}%
\frac{\partial ^{2}}{\partial \varphi ^{2}}\right.  \notag \\
&&+\left. \alpha \frac{1}{4}\left( \frac{a+2b\sin \theta }{ab+b^{2}\sin
\theta }\right) ^{2}-\beta \frac{\sin \theta }{ab+b^{2}\sin \theta }\right] ,
\label{h}
\end{eqnarray}%
where, 
\begin{equation*}
M=-\frac{1}{2}\frac{a+2b\sin \theta }{ab+b^{2}\sin \theta },\text{ }K=\frac{%
\sin \theta }{ab+b^{2}\sin \theta }.
\end{equation*}%
We are ready to construct commutator $[A,B]$ of two variables $A$ and $B$ in
quantum mechanics, which can be straightforwardly realized by a direct
correspondence of the Dirac's brackets as $\{A,B\}_{D}\rightarrow \left[ A,B%
\right] /i\hbar $. From the Dirac's brackets (\ref{xp1}), the first category
of the fundamental commutators between operators $q^{\mu }$ and $p_{\nu }$
are given by,%
\begin{equation}
\lbrack q^{\mu },q^{\nu }]=0,\text{ }[p_{\mu },p_{\nu }]=0,\text{ }[q^{\mu
},p_{\nu }]=i\hbar \delta _{\nu }^{\mu }.  \label{xp2}
\end{equation}%
In light of the GTCQ, we have the\ second category of fundamental
commutators between $q^{\mu }$ and $H$ from Eq. (\ref{xh}),%
\begin{eqnarray}
\left[ \theta ,H\right] &=&\frac{\hbar ^{2}}{mb^{2}}\left( \frac{\partial }{%
\partial \theta }+\frac{b\cos \theta }{2\left( a+b\sin \theta \right) }%
\right) =i\hbar \frac{p_{\theta }}{mb^{2}},  \label{qxh1} \\
\left[ \varphi ,H\right] &=&\frac{\hbar ^{2}}{m(a+b\sin \theta )^{2}}\frac{%
\partial }{\partial \varphi }=i\hbar \frac{p_{\varphi }}{m(a+b\sin \theta
)^{2}}.  \label{qxh2}
\end{eqnarray}%
From these quantum commutators, the operators $p_{\theta }$ and $p_{\varphi
} $ are, respectively,%
\begin{equation}
p_{\theta }=-i\hbar \left[ \frac{\partial }{\partial \theta }+\frac{b\cos
\theta }{2\left( a+b\sin \theta \right) }\right] ,\text{ }p_{\varphi
}=-i\hbar \frac{\partial }{\partial \varphi }\text{.}  \label{cmom}
\end{equation}%
Using these operators, we can directly calculate two quantum commutators $%
\left[ p_{\theta },H\right] $ and $\left[ p_{\varphi },H\right] $ with
quantum Hamiltonian (\ref{h}), and the results are, respectively,%
\begin{eqnarray}
\left[ p_{\theta },H\right] &=&i\hbar \frac{b\cos \theta }{m(a+b\sin \theta
)^{3}}p_{\varphi }^{2}+i\hbar \frac{\hbar ^{2}\cos \theta \left(
a^{2}(\alpha -2\beta +1)+2ab(\alpha -\beta )\sin \theta -b^{2}\right) }{%
4bm(a+b\sin \theta )^{3}},  \label{qph1} \\
\left[ p_{\varphi },H\right] &=&0.  \label{qph2}
\end{eqnarray}%
The second equation (\ref{qph2}) is satisfactory, whereas the first one (\ref%
{qph1})\ can hardly hold true. In the GTCQ, the\ quantum commutator $\left[
p_{\theta },H\right] $ (\ref{qph1}) must be the canonical quantization of
the Dirac bracket (\ref{ph}). We get, with noting the mutual commutabiliy
between two observables $p_{\varphi }$ and $\theta $, 
\begin{equation}
i\hbar \left\{ p_{\theta },H\right\} _{D}=\frac{i\hbar b\cos \theta
p_{\varphi }^{2}}{m(a+b\sin \theta )^{3}}.  \label{qph1-1}
\end{equation}%
In comparison with the right-handed sides of the Eqs. (\ref{qph1}) and (\ref%
{qph1-1}), we obtain a unique solution, 
\begin{equation}
\alpha =\beta =\frac{a^{2}-b^{2}}{a^{2}}(\neq 1),
\end{equation}%
which leads an unacceptable curvature dependent quantum potential that
includes the extrinsic curvature $M$,%
\begin{equation}
V_{D}=-\frac{\hbar ^{2}}{2m}\frac{a^{2}-b^{2}}{a^{2}}\left( M^{2}-K\right) =-%
\frac{\hbar ^{2}}{2m}\frac{a^{2}-b^{2}}{4b^{2}\left( a+b\sin \theta \right)
^{2}}.
\end{equation}%
However, no matter what other values of\ $\alpha $ and $\beta $ are chosen,
there is a manifest breakdown of the canonical quantization rule between
Dirac bracket\ $\left\{ p_{\theta },H\right\} _{D}$ (\ref{ph}) and the
quantum\ commutator $\left[ p_{\theta },H\right] $ (\ref{qph1}). So we see
that the intrinsic geometry is insufficient for the GTCQ to be
self-consistent.

If using original form of the Dirac's theory instead, we still have results (%
\ref{cmom})-(\ref{qph2}) but we can never require them as the canonical
quantization of the relevant Dirac brackets (\ref{xh})-(\ref{ph}). It is
sheer nonsense for we neither are able to exclude the extrinsic curvature $M$%
, nor give a unambiguous prediction of the curvature dependent potential to
be testable by experiment.

One should be noted that we have not introduced additional assumptions such
as "dummy factors" techniques \cite{Kleinert} etc. in passing from Dirac's
brackets to the quantum commutators. They mean further generalizations of
the Dirac's theory.

In classical limit $\hbar \rightarrow 0$, all inconsistency vanishes, as
expected.

\subsection{Summary}

From the studies in this section, we see that the GTCQ of second-class
constraints for quantum motion on the torus can not be consistently
formulated. We therefore need to invoke an extrinsic examination of the same
problem, as will be done in next section.

\section{GTCQ for a torus as a submanifold}

The surface equation of the torus (\ref{rr}) in Cartesian coordinates $%
\left( x,y,z\right) $ is given by,%
\begin{equation}
f\left( \mathbf{x}\right) \equiv a^{2}-b^{2}+(x^{2}+y^{2}+z^{2})-2a\sqrt{%
x^{2}+y^{2}}=0.
\end{equation}%
In this section, we will also first give the classical mechanics for motion
on the torus within the Dirac formalism of the classical mechanics for a
system with the second-class constraints, and then turn into quantum
mechanics. The GTCQ proves to be self-consistent all around and the
resultant momenta and Hamiltonian are exactly those given by the Eq. (\ref%
{gm}) and (\ref{gp}), respectively.

\subsection{Classical mechanical treatment}

The Lagrangian $L$ in the Cartesian coordinate system is,%
\begin{equation}
L=\frac{m}{2}\left( \dot{x}^{2}+\dot{y}^{2}+\dot{z}^{2}\right) -\lambda
f\left( \mathbf{x}\right) .  \label{lagca}
\end{equation}%
The generalized momentum $\mathbf{p}$ whose three components $p_{i}$ $%
(i=x,y,z)$ and $p_{\lambda }$ canonically conjugate to variables $x_{i}$ $%
(x_{1}=x,x_{2}=y,x_{3}=z,)$ and $\lambda $, are given by, respectively,%
\begin{eqnarray}
p_{i} &=&\frac{\partial L}{\partial \dot{x}_{i}}=m\dot{x}_{i},(i=1,2,3), \\
p_{\lambda } &=&\frac{\partial L}{\partial \dot{\lambda}}=0.  \label{plambca}
\end{eqnarray}%
Eq. (\ref{plambca}) represents the primary constraint,%
\begin{equation}
\varphi _{1}\equiv p_{\lambda }\approx 0.  \label{prim2}
\end{equation}%
By the Legendre transformation, the primary Hamiltonian $H_{p}$ is,%
\begin{equation}
H_{p}=\frac{1}{2m}p_{i}^{2}+\lambda f\left( \mathbf{x}\right) +\dot{\lambda}%
p_{\lambda }.
\end{equation}%
The secondary constraints are determined by\ successive use of the Poisson
brackets,%
\begin{eqnarray}
\varphi _{2} &\equiv &\left\{ \varphi _{1},H_{p}\right\}
=-(a^{2}-b^{2}+x_{i}^{2}-2a\sqrt{x^{2}+y^{2}})\approx 0,  \label{1st2} \\
\varphi _{3} &\equiv &\left\{ \varphi _{2},H_{p}\right\} =-\frac{2\left( 
\sqrt{x^{2}+y^{2}}(p_{x}x+p_{y}y+p_{z}z)-a(p_{x}x+p_{y}y)\right) }{m\sqrt{%
x^{2}+y^{2}}}\approx 0, \\
\varphi _{4} &\equiv &\left\{ \varphi _{3},H_{p}\right\} =\frac{4\lambda
\left( a^{2}-2a\sqrt{x^{2}+y^{2}}+x_{i}^{2}\right) }{m}+\frac{%
2a(p_{y}x-p_{x}y)^{2}}{m^{2}\left( x^{2}+y^{2}\right) ^{3/2}}-\frac{%
2p_{i}^{2}}{m^{2}}\approx 0,  \label{thica} \\
\varphi _{5} &\equiv &\left\{ \varphi _{4},H_{p}\right\} =\frac{4\dot{\lambda%
}\left( a^{2}-2a\sqrt{x^{2}+y^{2}}+x_{i}^{2}\right) }{m}-\frac{%
6a(p_{x}x+p_{y}y)(p_{y}x-p_{x}y)^{2}}{m^{3}\left( x^{2}+y^{2}\right) ^{5/2}}%
\approx 0.  \label{forca}
\end{eqnarray}%
Similarly, the Dirac bracket between two variables $A$ and $B$ is defined by,%
\begin{equation}
\left\{ A,B\right\} _{D}=\left\{ A,B\right\} -\left\{ A,\varphi _{u}\right\}
D_{uv}^{-1}\left\{ \varphi _{v},B\right\} ,
\end{equation}%
where the $4\times 4$ matrix $D\equiv \left\{ D_{uv}\right\} $ whose
elements are defined by $D_{uv}\equiv \left\{ \varphi _{u},\varphi
_{v}\right\} $ with $u,v=1,2,3,4$ from Eqs. (\ref{prim2}) and (\ref{1st2})-(%
\ref{thica}). The inverse matrix $D^{-1}$ is easily carried out,%
\begin{equation}
D^{-1}=\left( 
\begin{array}{cccc}
0 & D_{12}^{-1} & 0 & \kappa \\ 
-D_{12}^{-1} & 0 & -\kappa & 0 \\ 
0 & \kappa & 0 & 0 \\ 
-\kappa & 0 & 0 & 0%
\end{array}%
\right) ,
\end{equation}%
where,%
\begin{equation}
D_{12}^{-1}=\frac{\left( 3a^{2}-7a\sqrt{x^{2}+y^{2}}\right)
(p_{y}x-p_{x}y)^{2}+4\left( x^{2}+y^{2}\right) ^{2}p_{i}^{2}}{4b^{4}m\left(
x^{2}+y^{2}\right) ^{2}},\text{ }\kappa =\frac{m}{4b^{2}}.
\end{equation}%
Then primary Hamiltonian $H_{p}$ assumes its usual one: $H_{p}\rightarrow
H_{c},$%
\begin{equation}
H_{c}=\frac{p_{x}^{2}+p_{y}^{2}+p_{z}^{2}}{2m}.  \label{HP}
\end{equation}%
All fundamental Dirac's brackets are as follows,%
\begin{eqnarray}
\{x_{i},x_{j}\}_{D} &=&0,  \label{xxca1} \\
\{x_{i},p_{j}\}_{D} &=&\delta _{ij}-\frac{1}{b^{2}}f_{i}f_{j},  \label{xpca1}
\\
\{p_{i},p_{j}\}_{D} &=&-\frac{1}{b^{2}}\left[ f_{i}\left( p_{j}+\frac{%
a\left( xp_{y}-yp_{x}\right) }{\left( x^{2}+y^{2}\right) ^{3/2}}\left(
y\delta _{1j}-x\delta _{2j}\right) \right) -f_{j}\left( p_{i}+\frac{a\left(
xp_{y}-yp_{x}\right) }{\left( x^{2}+y^{2}\right) ^{3/2}}\left( y\delta
_{1i}-x\delta _{2i}\right) \right) \right] ,  \label{xhca1} \\
\{x_{i},H_{c}\}_{D} &=&\frac{p_{i}}{m}=\dot{x}_{i}, \\
\{p_{i},H_{c}\}_{D} &=&-\frac{1}{mb^{2}}\left[ f_{i}\left(
p_{x}^{2}+p_{y}^{2}+p_{z}^{2}-\frac{a\left( xp_{y}-yp_{x}\right) ^{2}}{%
\left( x^{2}+y^{2}\right) ^{3/2}}\right) \right] =\dot{p}_{i},  \label{phca1}
\end{eqnarray}%
\ where $f_{i}=x_{i}-a\left( x\delta _{1i}+y\delta _{2i}\right) /\sqrt{%
x^{2}+y^{2}}$.

\subsection{Quantum mechanical treatment}

Now let us turn to quantum mechanics. The first category of the fundamental
commutators between operators $x_{i}$ and $p_{i}$ are, by quantization of (%
\ref{xxca1})-(\ref{xhca1}),%
\begin{eqnarray}
\left[ x_{i},x_{j}\right] &=&0,\text{ \ }\left[ x_{i},p_{j}\right] =i\hbar
\left( \delta _{ij}-\frac{1}{b^{2}}f_{i}f_{j}\right) ,  \label{xx-xp} \\
\left[ p_{i},p_{j}\right] &=&-\frac{i\hbar }{b^{2}}\left[ f_{i}\left( p_{j}+a%
\frac{L_{z}\left( y\delta _{1j}-x\delta _{2j}\right) +\left( y\delta
_{1j}-x\delta _{2j}\right) L_{z}}{2\left( x^{2}+y^{2}\right) ^{3/2}}\right)
\right.  \notag \\
&&\left. -f_{j}\left( p_{i}+a\frac{L_{z}\left( y\delta _{1i}-x\delta
_{2i}\right) +\left( y\delta _{1i}-x\delta _{2i}\right) L_{z}}{2\left(
x^{2}+y^{2}\right) ^{3/2}}\right) \right] ,  \label{ppca2}
\end{eqnarray}%
where $L_{z}=xp_{y}-yp_{x}$. It seems that we have complicated
operator-ordering problem as passing from the Dirac bracket Eq. (\ref{xhca1}%
) to the quantum commutator (\ref{ppca2}). In fact, only one pair between
the noncommuting observables $x_{i}$ (precisely, $\left( y\delta
_{1j}-x\delta _{2j}\right) $) and $L_{z}$ matters, and the product of $%
\left( y\delta _{1j}-x\delta _{2j}\right) $ and $L_{z}$ can be made
Hermitian by a symmetric construction, $\left( \left( y\delta _{1j}-x\delta
_{2j}\right) L_{z}+L_{z}\left( y\delta _{1j}-x\delta _{2j}\right) \right) /2$%
. Other products of factors $f_{i}$ (or $f_{j}$) and $L_{z}$ impose no
operator-ordering problem because of the Jacobi identity.

There is a family of the momenta $p_{i}$ all of them are solutions to the
Eq. (\ref{ppca2}), as explicitly shown in \cite{japan1992}. With these
momenta $p_{i}$ at hand, we completely do not know the correct form of the
quantum Hamiltonian, as suggested by Eq. (\ref{HP}). It is therefore
understandable that the quantum Hamiltonian would contain arbitrary
parameters.

However, the GTCQ requires the second category of the fundamental
commutators as $\left[ x_{i},H\right] $ and $\left[ p_{i},H\right] $. We
immediately find that the momenta $p_{i}$ from following commutators, 
\begin{equation}
\left[ x_{i},H\right] =i\hbar \frac{p_{i}}{m}.  \label{xhca2}
\end{equation}%
The obtained momenta $p_{i}$ are nothing but three components of the
geometric momentum (\ref{gm}) on the torus \cite{torus4}, 
\begin{eqnarray}
p_{x} &=&-i\hbar \left( \frac{\cos \theta \cos \varphi }{b}\frac{\partial }{%
\partial \theta }-\frac{\sin \varphi }{a+b\sin \theta }\frac{\partial }{%
\partial \varphi }-\frac{a+2b\sin \theta }{2b(a+b\sin \theta )}\sin \theta
\cos \varphi \right) , \\
p_{y} &=&-i\hbar \left( \frac{\cos \theta \sin \varphi }{b}\frac{\partial }{%
\partial \theta }+\frac{\cos \varphi }{a+b\sin \theta }\frac{\partial }{%
\partial \varphi }-\frac{a+2b\sin \theta }{2b(a+b\sin \theta )}\sin \theta
\sin \varphi \right) , \\
p_{z} &=&i\hbar \left( \frac{\sin \theta }{b}\frac{\partial }{\partial
\theta }+\frac{a+2b\sin \theta }{2b\left( a+b\sin \theta \right) }\cos
\theta \right) .
\end{eqnarray}

As to the form of quantum Hamiltonian, we also start from the general form (%
\ref{h}), and now resort to the following complicated operator-ordering
arrangement with $w_{\pm }=(x\pm iy)^{3/2}$, 
\begin{eqnarray}
\left[ p_{i},H\right] &=&-\frac{i\hbar }{mb^{2}}\{mHf_{i}+mf_{i}H  \notag \\
&&-\frac{a}{4}\alpha _{1}[f_{i}(L_{z}\frac{1}{w_{+}}L_{z}\frac{1}{w_{-}}+%
\frac{1}{w_{-}}L_{z}\frac{1}{w_{+}}L_{z})+(L_{z}\frac{1}{w_{+}}L_{z}\frac{1}{%
w_{-}}+\frac{1}{w_{-}}L_{z}\frac{1}{w_{+}}L_{z})f_{i}]  \notag \\
&&-\frac{a}{4}\alpha _{2}[f_{i}(L_{z}\frac{1}{w_{+}}\frac{1}{w_{-}}L_{z}+%
\frac{1}{w_{-}}L_{z}L_{z}\frac{1}{w_{+}})+(L_{z}\frac{1}{w_{+}}\frac{1}{w_{-}%
}L_{z}+\frac{1}{w_{-}}L_{z}L_{z}\frac{1}{w_{+}})f_{i}]  \notag \\
&&-\frac{a}{4}\alpha _{3}[f_{i}(\frac{1}{w_{+}}L_{z}L_{z}\frac{1}{w_{-}}%
+L_{z}\frac{1}{w_{-}}\frac{1}{w_{+}}L_{z})+(\frac{1}{w_{+}}L_{z}L_{z}\frac{1%
}{w_{-}}+L_{z}\frac{1}{w_{-}}\frac{1}{w_{+}}L_{z})f_{i}]  \notag \\
&&-\frac{a}{4}\alpha _{4}[f_{i}(\frac{1}{w_{+}}L_{z}\frac{1}{w_{-}}%
L_{z}+L_{z}\frac{1}{w_{-}}L_{z}\frac{1}{w_{+}})+(\frac{1}{w_{+}}L_{z}\frac{1%
}{w_{-}}L_{z}+L_{z}\frac{1}{w_{-}}L_{z}\frac{1}{w_{+}})f_{i}]  \notag \\
&&-\frac{a}{2}\alpha _{5}\frac{1}{w_{+}w_{-}}(f_{i}L_{z}^{2}+L_{z}^{2}f_{i})%
\},  \label{op-ord}
\end{eqnarray}%
where $\alpha _{k}$, $(k=1,2,...5)$ are five real parameters satisfying $%
\sum \alpha _{k}=1$. In comparison of both sides of the this equation, we
find that the solution $\alpha =\beta =1$, and two of the five real
parameters $\alpha _{k}$ are freely to be specified, 
\begin{equation}
\alpha _{1}=\frac{11}{9}-\alpha _{4}-\alpha _{5},\alpha _{2}=\alpha _{3}=-%
\frac{1}{9}.
\end{equation}%
We see that free parameters remain, but they are irrelevant to observable
quantities such as momentum and potential. In fact, with $\alpha =\beta =1$
in (\ref{vd}), a much simpler choice of the operator-ordering without free
parameters is possible,%
\begin{eqnarray}
\left[ p_{i},H\right] &=&-\frac{i\hbar }{mb^{2}}\{mHf_{i}+mf_{i}H+\frac{1}{9}%
\frac{a}{4}[(\frac{1}{w_{+}}f_{i}L_{z}^{2}\frac{1}{w_{-}}+\frac{1}{w_{-}}%
f_{i}L_{z}^{2}\frac{1}{w_{+}})  \notag \\
&&+(\frac{1}{w_{+}}L_{z}^{2}\frac{1}{w_{-}}f_{i}+\frac{1}{w_{-}}L_{z}^{2}%
\frac{1}{w_{+}}f_{i})]-\frac{10}{9}\frac{a}{2}\frac{1}{w_{+}w_{-}}%
(f_{i}L_{z}^{2}+L_{z}^{2}f_{i})\}.  \label{ph2}
\end{eqnarray}%
Even we can by no means exhaust all possible forms of the operator-ordering,
from Eqs. (\ref{op-ord}) and (\ref{ph2}), we can at least conclude that the
curvature dependent potential (\ref{vd}) given by the Dirac formalism
converges to the geometric potential (\ref{gp}) given by the Schr\"{o}dinger
one.

\subsection{Summary}

An examination of the motion on torus as\ a submanifold problem in GTCQ
ensures a highly self-consistent description, and this formalism comes
compatible with the Schr\"{o}dinger one.

\section{Remarks and conclusions}

It is long known that Dirac's theory of second-class constraints, in which
the fundamental commutation relations involve only those between canonical
positions\ and canonical momenta, contains redundant freedoms and causes
difficulty sometimes.\ To overcome the problems, we recently put forward a
proposal that the commutators between the positions, momenta, and
Hamiltonian form a full set of the fundamental commutation relations to
construct a self-consistent quantum theory, the so-called GTCQ. Then the
GTCQ produces a unique form of the geometric momentum, and imposes
additional requirement on the form of the Hamiltonian via the curvature
dependent potential that has no direct analogy. We see that the geometric
potential comes as the consequence of the extrinsic examination of the
constrained motion.

Through a careful analysis of the quantum motion on a torus, we demonstrate
that the purely intrinsic geometry does not suffice for the GTCQ to be
self-consistently formulated, but an extrinsic examination of the torus in
three dimensional flat space does. Our study implies that the Dirac
formalism is complementary to the Schr\"{o}dinger one. The former can be
helpful to eliminate the intrinsic description, and the latter gives the
unique form of the geometric potential, while both define the identical form
of the geometric momentum.

\begin{acknowledgments}
This work is financially supported by National Natural Science Foundation of
China under Grant No. 11175063.
\end{acknowledgments}


\begin{thebibliography}{99}
\bibitem{dirac3} P. A. M. Dirac, \textit{Proc. R. Soc. Lond. A} \textbf{109}%
, 642-653(1925). In this paper, Dirac scrutinized the problem whether $%
i\hbar \{f,H\}\rightarrow \lbrack f,H]$ is necessary for a system without
constraints and wrote his thoughtful observations as follows
\textquotedblleft \textit{These equations (i.e. }$i\hbar \{f,H\}\rightarrow
\lbrack f,H]$\textit{) will be true on the quantum theory for systems for
which the orders of the factors of the products occurring in the equations
of motions are unimportant. They may be taken to be true for system for
which these orders are important in one can decide upon the orders of the
factors in }$H$\textit{.}\textquotedblright
\end{thebibliography}
\end{document}